\documentclass[11pt]{article}

\usepackage[a4paper,margin=1in]{geometry}
\usepackage{graphicx}
\usepackage{hyperref}
\usepackage{authblk}
\usepackage{setspace}
\usepackage{enumitem}
\usepackage{caption}
\usepackage{booktabs}
\usepackage{orcidlink}
\usepackage{cleveref}
\hypersetup{
  colorlinks=true,
  linkcolor=blue,
  citecolor=blue,
  urlcolor=blue,
  pdftitle={Fetpype: An Open-Source Pipeline for Reproducible Fetal Brain MRI Analysis}
}

\title{\textbf{Fetpype: An Open-Source Pipeline for Reproducible Fetal Brain MRI Analysis}}

\author[1,2,*]{Thomas Sanchez\orcidlink{0000-0003-3668-5155}}
\author[3,*]{Gerard Martí-Juan\orcidlink{0000-0003-4729-7182}}
\author[4]{David Meunier\orcidlink{0000-0002-5812-6138}}
\author[3,5]{Miguel Angel Gonzalez Ballester\orcidlink{0000-0002-9227-6826}}
\author[3]{Oscar Camara\orcidlink{0000-0002-5125-6132}}
\author[6,7]{Elisenda Eixarch\orcidlink{0000-0001-7379-9608}}
\author[3]{Gemma Piella\orcidlink{0000-0001-5236-5819}}
\author[1,2]{Meritxell Bach Cuadra\orcidlink{0000-0003-2730-4285}}
\author[4]{Guillaume Auzias\orcidlink{0000-0002-0414-5691}}

\affil[1]{CIBM – Center for Biomedical Imaging, Switzerland}
\affil[2]{Department of Diagnostic and Interventional Radiology, Lausanne University\newline Hospital and University of Lausanne, Switzerland}
\affil[3]{BCN MedTech, Department of Engineering, Universitat Pompeu Fabra, Spain}
\affil[4]{Aix-Marseille Université, CNRS, Institut de Neurosciences de La Timone, France}
\affil[5]{ICREA, Barcelona, Spain}
\affil[6]{BCNatal | Fetal Medicine Research Center (Hospital Clínic and Hospital Sant Joan de Déu, Universitat de Barcelona), Barcelona, Spain}
\affil[7]{Institut d’Investigacions Biomèdiques August Pi i Sunyer (IDIBAPS), Barcelona, Spain and Centre for Biomedical Research on Rare Diseases (CIBERER), Barcelona, Spain}
\affil[*]{Equal contribution}

\date{}

\begin{document}

\maketitle

\section{Summary}

Fetal brain magnetic resonance imaging (MRI) is crucial for assessing neurodevelopment \textit{in utero}. However, fetal MRI analysis remains technically challenging due to fetal motion, low signal-to-noise ratio, and the need for complex multi-step processing pipelines. These pipelines typically include motion correction, super-resolution reconstruction, tissue segmentation, and cortical surface extraction. While specialized tools exist for each individual processing step, integrating them into a robust, reproducible, and user-friendly end-to-end workflow remains difficult. This fragmentation limits reproducibility across studies and hinders the adoption of advanced fetal neuroimaging methods in both research and clinical contexts.

\texttt{Fetpype} addresses this gap by providing a standardized, modular, and reproducible framework for fetal brain MRI preprocessing and analysis, enabling researchers to process raw T2-weighted acquisitions through to derived volumetric and surface-based outputs within a unified workflow.

\section{Statement of need}
\texttt{Fetpype} is an open-source Python package designed to streamline and standardize the preprocessing and analysis of T2-weighted fetal brain MRI data. The package targets the fetal neuroimaging community, where methodological heterogeneity and complex software dependencies have historically limited reproducibility and comparability across studies.

Existing fetal brain MRI tools typically focus on individual processing steps and require customized code for pre- and post-processing, as well as to connect different modules, making it difficult to reproduce processing results across studies. \texttt{Fetpype} addresses these challenges by providing a configurable, containerized, and \texttt{Nipype}-driven solution that integrates state-of-the-art fetal MRI processing tools into a cohesive pipeline. By emphasizing reproducibility, extensibility, and ease of use, \texttt{Fetpype} lowers the barrier to applying advanced fetal MRI analysis methods and facilitates consistent processing across sites, scanners, and studies.  In doing so, \texttt{Fetpype} improves comparability across studies and supports community collaboration by facilitating the dissemination of new image processing methods for clinical applications. The pipeline is publicly available \href{https://github.com/fetpype/fetpype}{on GitHub}.

\begin{figure}[tbp]
    \centering
    \includegraphics[width=\linewidth]{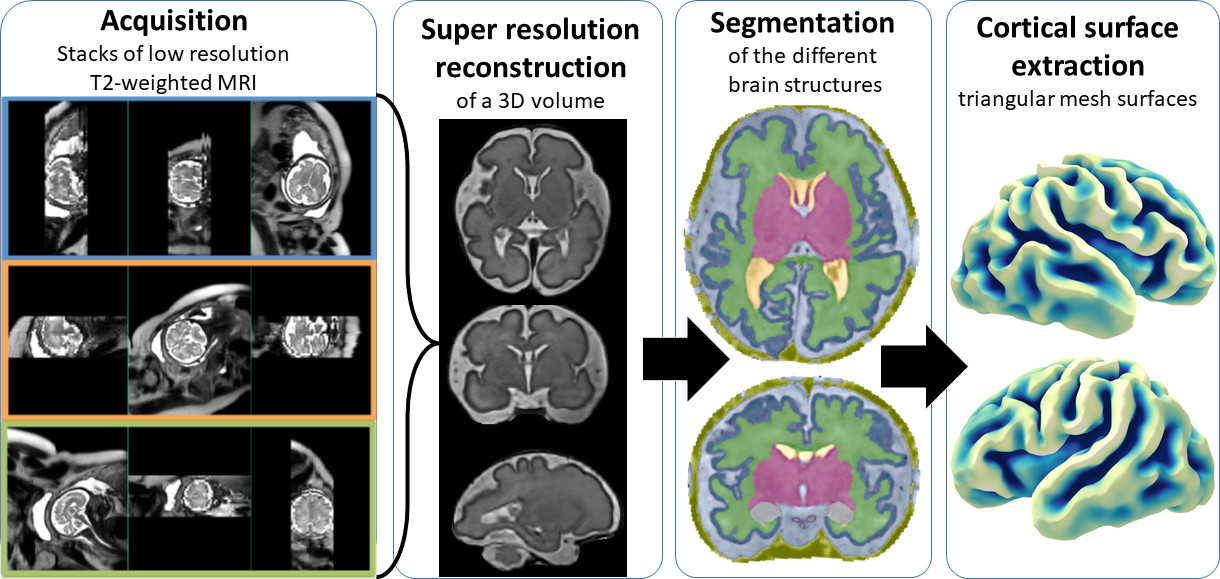}
    \caption{The different steps covered by \texttt{Fetpype}. Starting from several T2-weighted stacks of thick slices of the fetal brain (\textit{acquisition}), \texttt{Fetpype} pre-processes data before feeding them to a \textit{super-resolution reconstruction} algorithm that fuses them into a single high-resolution volume. This volume then undergoes \textit{segmentation}, before moving to cortical \textit{surface extraction}.}
    \label{fig:fetpype}
\end{figure}

\section{State of the field}
Fetal brain MRI analysis relies on a range of specialized tools that address individual processing steps, particularly motion correction and super-resolution reconstruction. Widely used reconstruction frameworks include SVRTK (C++) \cite{kuklisova-murgasova_reconstruction_2012,uus2022automated}, NiftyMIC (Python) \cite{ebner2020automated}, and NeSVoR (Python/PyTorch) \cite{xu2023nesvor}. While increasingly distributed as containers, these tools rely on distinct data organization schemes as well as custom pre- and post-processing steps. Downstream processing tools for brain extraction, segmentation, and surface reconstruction show similar diversity, combining Python scripts, compiled binaries, and domain-specific imaging libraries \cite{makropoulos2018developing,uus2023bounti,faghihpirayesh2024fetal,bazin2005topology}. As a result, constructing an end-to-end fetal MRI workflow typically requires custom scripting to orchestrate multiple containers, manage data formats, and handle intermediate outputs, limiting reproducibility and accessibility.

\texttt{Fetpype} was built to address these limitations by providing a unified, Python-based framework to integrate existing containerized tools for fetal brain MRI analysis. By enforcing data formatting following the widely-used Brain Imaging Data Structure (BIDS) standard \cite{gorgolewski2016brain}, and leveraging containerized execution and \texttt{Nipype}-based workflow management \cite{gorgolewski2011nipype}and Hydra-based configuration \cite{Yadan2019Hydra}, \texttt{Fetpype} provides a standardized and scalable environment for reproducible end-to-end fetal brain MRI analysis on both local workstations and large-scale computing clusters.

\section{Software design}
\texttt{Fetpype} is built around four core design principles: data standardization, containerization, workflow orchestration, and flexible configuration:

\begin{enumerate}[label=\arabic*.]
    \item \textbf{Data Standardization}: \texttt{Fetpype} expects input data organized according to the Brain Imaging Data Structure (BIDS) standard \cite{gorgolewski2016brain}, promoting interoperability and simplifying data management.
    \item \textbf{Containerization}: Individual processing tools are encapsulated within Docker or Singularity containers. This ensures reproducibility and reduces installation issues, providing a better experience for the end user.
    \item \textbf{Workflow Management}: The \texttt{Nipype} library \cite{gorgolewski2011nipype} is used to construct processing workflows. It provides a robust interface for combining different steps from different containers or packages, facilitating data caching and parallelization, and allowing pipelines to be easily shareable.
    \item \textbf{Configuration}: Pipeline configuration is managed using simple YAML files and the Hydra library \cite{Yadan2019Hydra}, allowing users to easily select between different modules or parameters without directly modifying the code.
\end{enumerate}

The current implementation of \texttt{Fetpype} integrates modules for:
\begin{itemize}
    \item \textbf{Data preprocessing}: including brain extraction using \texttt{Fetal-BET} \cite{faghihpirayesh2024fetal}, non-local means denoising \cite{manjon2010adaptive}, and N4 bias-field correction \cite{tustison2010n4itk}, all wrapped into a single container built at \url{https://github.com/fetpype/utils_container}.
    \item \textbf{Super-resolution reconstruction}: implementing three widely used pipelines: NeSVoR \cite{xu2023nesvor}, SVRTK \cite{kuklisova-murgasova_reconstruction_2012,uus2022automated}, and NiftyMIC \cite{ebner2020automated}.
    \item \textbf{Segmentation}: implementing BOUNTI \cite{uus2023bounti} and the Developing Human Connectome Project pipeline \cite{makropoulos2018developing}.
    \item \textbf{Cortical surface extraction}: using a custom implementation available at \url{https://github.com/fetpype/surface_processing} based on \cite{bazin2005topology,bazin2007topology,ma2022cortexode}.
\end{itemize}

The overall processing workflow is summarized in Figure~\ref{fig:fetpype}.

\section{Research impact statement}
\texttt{Fetpype} is the result of a longstanding collaboration within a European consortium of researchers specializing in fetal neuroimaging. Its default configurations and processing workflows have been the result of extensive testing to achieve robust processing on data acquired across multiple hospitals in France, Spain, and Switzerland, covering a range of scanners and acquisition protocols.

The framework has been used to process large-scale fetal MRI datasets within the consortium, has contributed to a first publication \cite{sanchez2026data}, and is supporting ongoing research projects. \texttt{Fetpype} is used by multiple research groups and has begun to receive external contributions, including pull requests that integrate additional processing methods. This suggests that \texttt{Fetpype} addresses a clear methodological need and can serve as shared community infrastructure for fetal brain MRI research.

In the future, we plan to supplement \texttt{Fetpype} with an automated reporting library containing automated quality control \cite{sanchez2025automatic}, subject-wise and population-wise biometry and volumetry \cite{esteban2017mriqc,neves2025scanner}, as well as spectral analysis of surfaces \cite{germanaud2012larger}. We welcome community contributions, particularly implementations of new methods that can be integrated into the existing containerized workflow framework.

\section*{AI usage disclosure}
GitHub Copilot, integrated within Visual Studio Code, was used during software development to assist with code completion and implementation. ChatGPT (GPT-5.2) was used for proofreading and language refinement of the manuscript. The authors take full responsibility for the written content.

\section*{Acknowledgements}
This work was funded by Era-net NEURON MULTIFACT project (TS: Swiss National Science Foundation grants 31NE30\_203977, 215641; GA: French National Research Agency, Grant ANR-21-NEU2-0005; EE: Instituto de Salud Carlos III (ISCIII) grant AC21\_2/00016; GMJ, GP, OC, MAGB:  Ministry of Science, Innovation and Universities: MCIN/AEI/10.13039/501100011033/), the SulcalGRIDS Project, (GA: French National Research Agency Grant ANR-19-CE45-0014), the pediatric domain shifts project (TS: SNSF 205320-215641), and NVIDIA research grants with the use of NVIDIA RTX6000 ADA GPUs.  We acknowledge the CIBM Center for Biomedical Imaging, a Swiss research center of excellence founded and supported by CHUV, UNIL, EPFL, UNIGE and HUG.

\bibliographystyle{plain}
\bibliography{paper}

\end{document}